\documentstyle[12pt]{article}

\begin{document}
\renewcommand{\thefootnote}{\fnsymbol{footnote}}

\newpage
\setcounter{page}{0}


%

\begin{center}
{\LARGE {\bf The Gel'fand-Tsetlin Selection Rules and \\ [0.3cm]
  Representations of Quantum Algebras }}\\ [1cm] 
{\large  A.N. Leznov$^{a,b}$ }\\
{\em  Instituto de Investigaciones en Matem\'aticas
  Aplicadas y en Sistemas}~\\
{\em Universidad Nacional Aut\`onoma de M\`exico} \\
{\em Apartado Postal 48-3, 62251 Cuernavaca,    \\
     Morelos, M\'exico} \\
\quad \\
{\em {~$~^{(a)}$ On leave from Institute for High Energy Physics,}}\\
{\em 142284 Protvino, Moscow Region, Russia}\\
{\em {~$~^{(b)}$ Also at Bogoliubov Laboratory of Theoretical Physics,
JINR,}}\\
{\em 141980 Dubna, Moscow Region, Russia}~\quad\\

\end{center}

\vfill

\centerline{ {\bf Abstract}}

The problem of construction of irreducible representations of quantum
$A^q_n$ algebras is solved at the level of explicit integration of the
linear (inhomogeneous) system in finite differences in the
n-dimensional space. The general solution of this system is  given
explicitly and particular ones, which correspond to the irreducible
representations are selected.

\vfill
{\em e-mail:\\
1) leznov@ce.ifisicam.unam.mx }

\newpage

\section{Introduction}

In the recent paper of the author \cite{Leznov} the problem of construction
of irreducible representations of the unimodular matrix algebra $A_n$ -- the
famous result of Gel'fand and Tsetlin \cite{Tsetlin} -- was reformulated  at
the level of explicit integration of a linear system in finite differences.

It is possible to reduce all of the calculations to the solution of a
linear system of equations in finite differences (or PDE's in  the
continuous functional group limit).
What is surprising and intriguing in this approach is the
fact that the results obtained at the functional group level \cite{Eisenhart}
\cite{Malkin} (Poisson brackets of the classical Hamiltonian formalism)
literally pass to those arising in the algebraic calculations. The only thing
which is necessary is to replace Poisson brackets by commutators, and
canonically conjugated classical variables by elements of n-dimensional
Heisenberg algebras and to choose the correct order of operators involved. 
All expressions conserve their form at both levels after
replacing the operation of usual differentiation in the case of the
functional group by discrete difference in the algebra-theoretical
case.

In the present paper we apply the method of Ref.\ \cite{Leznov}  to the
problem of construction of irreducible representations of the quantum $A^q_n$
algebras.
We try to resolve the system of commutation relations between $3n$ generators
($2n$ generators of the simple roots $X^{\pm}_i$ and $n$ generators of the
Cartan elements $h_i$)
\begin{equation}
[X^+_i,X^-_j]=\delta_{ij} \phi_i(h_i),\quad [h_i,X^{\pm}_j]=\pm k_{ji}X^{\pm}
_j \label{I}
\end{equation}
where $\phi_i$ are initially arbitrary functions of their arguments.

Using the "selection rules" of the Gel'fand-Tsetlin paper
\cite{Tsetlin} only, we find that
in the case of $U_q(2),\; A^q_1$ algebra it is possible to resolve in a
self-consistent way Eqs.\ (\ref{I}) for an arbitrary function $\phi$.
However, already in the case of $U_q(3),\, A^q_2$ algebras
the self-consistence of the construction leads to a
unique choice of the functions $\phi_i$
in the standard trigonometrical form.

The notion of co-product was not used in any step of our construction.
We try to conserve literally the text and
presentation of the material as in Ref.\ \cite{Leznov}, in order to simplify
for the reader the comparison between these too very similar papers.

In sections 2 and 3 we consider the simplest cases of $U_q(2),\,U_q(3)$
quantum algebras with detailized calculations. In section 4 we consider
the case of arbitrary $U_q(n+1)$ quantum algebras. In section 5 we summarize
the results and discuss the perspectives of the further investigations.

\section{The cases of $A^q_1 \approx SU_q(2)$ and $SU_q(2)$}

Every section will be divided into two parts: the
classical case (the functional group level)
and proper algebraic construction (''quantum case'').
As it was remarked in the introduction,
the classical results may be used as good hints in
further algebraic (quantum) calculations.

\subsection{The functional algebra case}

The functional $U_q(2)$ algebra contains four elements
$X^{\pm},H,I$, connected among themselves
by Poisson brackets:
\begin{equation}
 \{ I,X^{\pm}\}=\{I,H\}=0,\quad \{X^+,X^-\}=\phi (H),
 \quad \{H,X^{\pm}\}= \pm 2 X^{\pm} \label{2}
\end{equation}
where $\phi$ is an arbitrary function of a single argument.

In correspondence with the Darboux theorem \cite{Eisenhart} out of
the elements of the $U_q(2)$ functional group (\cite{Eisenhart})
it is possible to construct a pair of canonically conjugated variables
$M,m,\; \{M,m\}=1$ and two cyclic variables  $I,L$
which have zero Poisson brackets among themselves and with all
other elements of the $U_q(2)$ functional group (clearly, up to
an arbitrary canonical transformation).
The explicit expression for them in terms of
the functional group elements will be give a bit later.

Let us choose $M=H$ and $m={1\over 4} \ln {X^+ \over X^-}+f(H)$.
With the help of Poisson brackets (\ref{2}) it is not difficult
to see that thus constructed $M,m$ are really canonically
conjugated variables.
Resolving these relations with respect to the functional group elements
leads to the following realization
in terms of the canonical conjugated coordinate $m$ and momentum $M$
and the two cyclic momenta $L_1,L_2$:
\begin{eqnarray}
 X^+&=&{1\over 2}e^{2m} (f({L_1-L_2\over 2})-f(H)),
     \quad X^-={1\over 2} e^{-2m} (f({L_1-L_2\over 2})+f(H)),\nonumber \\
    H&=&M-{L_1+L_2\over 2}   ,\quad \phi(x)=
    \frac{d f^2}{dx}  \label{3}
\end{eqnarray}
It is not difficult to verify by the direct calculations that (\ref{3})
is indeed a realization of the functional group (\ref{2}).
If we want to restrict ourselves by the case of quantum $A^q_1$ algebra
it is necessary to put $I=0$.

\subsection{Quantum algebra case}

As always, to pass from the classical expressions to the quantum ones
it is necessary to order in some way the operators involved
and replace the Poisson brackets by commutators.
The equations (\ref{3}) tell us about very tempting
possibility to rewrite them as
\begin{eqnarray}
  X^+&=&{1\over 2}e^ m (f({L_1-L_2\over 2})-f(H)) e^ m,\quad X^-=
  {1\over 2}e^{-m} (f({L_1-L_2\over 2})+f(H)) e^{-m}, \label{4}      \\
  H&=&M-{L_1+L_2\over 2},\quad I=L_1+L_2,\quad L=X^+X^-+X^-X^++{1\over 4}
  (f^2(H+1)+f^2(H-1))  \nonumber
\end{eqnarray}
and consider now $M,m$ as generators of the Heisenberg algebra
($[M,m]=1,\;[M,1]=0,\;[m,1]=0$), with $L_1,L_2$ commuting with
all of the generators involved in (\ref{4}).
 Keeping in mind the following operator relation of Heisenberg algebra,
$\exp (\pm x) p \exp (\mp x)=p\mp 1$, we conclude that the generators
defined in (\ref{4}) satisfy the commutation relations (\ref{2})
of $U_q(2)$ algebra (of course with square brackets instead of
the curly ones) and $\phi={1\over 4}(f^2(H+1)-f^2(H-1))$.

Two Cazimir operators under realization (\ref{4}) take the constant values
\begin{equation}
 K^{(1}=L_1+L_2, \quad K^{(2}=X^+X^-+ X^-X^+ +{1\over 4}(f^2(H+1)+f^2(H-1))=
 {1\over 2}f^2\left({L_1-L_2\over 2}\right) \label{5}
\end{equation}
which proves the irreducibility of the constructed representation.
In the conclusion of this section we want to stress the fact the that
commutation   relations   by   themselves   give   no   additional
restrictions on the
form of the $f,\phi$ functions.

\section{$A^q_2 \equiv SU_q(3)$ and $U_q(3)$ cases}

In this case the problem consists in resolution of the system of commutation
relations
\begin{eqnarray}
 \left[X^+_1,X^-_1\right]=\phi(h_1),\quad \left[X^+_1,X^-_2\right]=0\quad
 \left[h_1,X^{\pm}_1\right]=\pm 2X^{\pm}_1,
\quad \left[h_1,X^{\pm}_2\right]=\mp X^{\pm}_2 \nonumber \\
 \left[X^+_1,X^-_2\right]=0,\quad \left[X^+_2,X^-_2\right]=\Phi(h_2)\quad
 \left[h_2,X^{\pm}_2\right]=\pm 2X^{\pm}_2,
\quad \left[h_2,X^{\pm}_1\right]=\mp X^{\pm}_1 . \label{6}
\end{eqnarray}
We choose the arbitrary functions $\phi(h_1),\Phi(h_2)$ in
correspondence with the comments in the end of the previous section.

The selection rules of GZ paper allow us to try to find a resolution of
this problem in the following form:
\begin{eqnarray}
 X^+_1 & =& {1\over 2}e^ m (f({L_1-L_2\over 2})-f(H)) e^ m,\quad
 X^+_2=e^{l_1}f^1e^{l_1}+e^{l_2} f^2 e^{l_2},\quad \nonumber \\
 h_1 &=& M-{L_1+L_2\over 2} \nonumber \\
  X^-_1 &=& {1\over 2}e^{-m} (f({L_1-L_2\over 2})+f(H)) e^{-m},\quad
  X^-_2=e^{-l_1} \bar f^1 e^{-l_1}+e^{-l_2} \bar f^2 e^{-l_2},\nonumber \\
 h_2 &=& -{M\over 2}+L_1+L_2- {N_1+N_2+N_3\over 2},     \label{7}
\end{eqnarray}
where all "structural" functions $f^{1,2},\bar f^{1,2}$ depend only on
momentum (capital letters) variables. We intentionally preserve the order of
factors to avoid rewriting the same formulae for several times.

\subsection{Functional group case}

In this case it is necessary to understand all of the above relations
at the functional group level. The commutators have to be replaced by
the Poisson brackets understood as usually:
$$
\{A,B\}=\sum^3_1 \left(\frac {\partial A}{\partial p_i}
  \frac {\partial B}{\partial x_i}-
  \frac {\partial A}{\partial x_i}\frac {\partial B}{\partial
   p_i}\right),
   \quad x_i=(m,l_1,l_2),\quad p_i=(M,L_1,L_2)
$$
Now all of the objects are commutative and the order of the factors
in (\ref{7}) is unimportant.
As a consequence of the vanishing Poisson brackets
$\{X^+_1,X^-_2\}=\{X^+_2, X^-_1\}=0$
we obtain equations determining the explicit dependence of structural
functions on momentum $M$.
Namely
$$
(\ln \bar f^1)_M+{1\over 2}{f'_1+f'_2\over f_1-f_2}=0,\quad
(\ln \bar f^2)_M+{1\over 2}{-f'_1+f'_2\over f_1-f_2}=0
$$
$$
(\ln f^1)_M+{1\over 2}{f'_1-f'_2\over f_1+f_2}=0 \quad
(\ln f^2)_M-{1\over 2}{f'_1+f'_2\over f_1+f_2}=0
$$
where $f_1\equiv f({L_1-L_2\over 2}),f_2\equiv f(M-{L_1+L_2\over 2})$
and sign ${}'$ means differentiation of the functions $f_1,f_2$
with respect to their arguments.

Poisson bracket $\{X^+_2,X^-_2\}=\Phi(h_2)$ have as its corollary three
additional equations:
$$
(\ln \bar f^1)_{L_2}+(\ln f^2)_{L_1}=0,\quad (\ln \bar f^2)_{L_1}+(\ln f^1)_
{L_2}=0
$$
\begin{equation}
-2(f^1\bar f^1)_{L_1}+2(f^2\bar f^2)_{L_2}=\Phi(h_2) \label{Z}
\end{equation}
The condition of  selfconsistency  of  second  pair  of  equations
containing
$\ln f, \ln \bar f$ with the first one  leads to the functional
equation with shifted arguments for a single unknown function
($X\equiv f(x), Y\equiv f(y)$):
\begin{equation}
(X_{xx}X+Y_{yy}Y)(X^2-Y^2)=(X_x^2-Y_y^2)(X^2+Y^2)\label{ZZ}
\end{equation}
This equation allows the exact integration and its general solution has the
form (see Appendix I)
\begin{equation}
f(x)=\sqrt {(a+b\cosh (2\epsilon x))} \label{VI}
\end{equation}

In what follows we will use a special choice of parameters
(\ref{VI}) such that $f=\sinh \epsilon x$. This choice can be
connected with the condition of the correct classical limit
(in the sense $\epsilon \to 0$) in which the quantum functional
algebra must pass to functional algebra $A_1$. In this case
it is possible to perform all the calculations
up to the explicit expressions.

Resolution of all additional equations except of (\ref{Z}) gives the
following dependence of the structure functions upon the momentum $M$:
$$
f^1=\cosh \epsilon {L_1-M\over 2} f^1(L_1,L_2),\quad f^2=\sinh \epsilon {M-
L_2\over 2} f^2(L_1,L_2)
$$
\begin{equation}
{}\label{SF}
\end{equation}
$$
\bar f^1=\sinh \epsilon {L_1-M\over 2} \bar f^1(L_1,L_2),\quad
\bar f^2=\cosh \epsilon {M-L_2\over 2} \bar f^2(L_1,L_2)
$$
where all new structure functions (for which we reserve the same notations)
depend only on two arguments $x_1\equiv L_1,x_2\equiv L_2$.

For two unknown functions $X^1\equiv -f^1\bar f^1,X^2\equiv f^2\bar f^2$
the equation (\ref{Z}) (keeping in mind the form of its right-hand side)
takes the form of the system of two equations:
\begin{equation}
(e^{\epsilon x_1} X^1)_{x_1}+(e^{\epsilon x_2} X^2)_{x_2}=e^{\epsilon(2x_1+2x
_2-N)}\quad(e^{-\epsilon x_1} X^1)_{x_1}+(e^{-\epsilon x_2} X^2)_{x_2}=e^
{-\epsilon (2x_1+2x_2-N)}\label{ZZZ}
\end{equation}
where $N\equiv N_1+N_2+N_3$.

In accordance with Ref.\ \cite{Leznov} we try to find a solution to the last
system in the form:
$$
X^1=Y^2_{x_2},\quad X^2=Y^1_{x_1}
$$
Solving the linear algebraic system of equations which arises for
the functions $Y^{1,2}$, we are led to the following solution of
the initial system:
$$
X^1={1\over 4\epsilon^2}({\sinh \epsilon (2x_1+x_2-N)\over \sinh \epsilon
(x_1-x_2)})_{x_2}+(\coth \epsilon (x_1-x_2)\Theta(x_1)+{\bar \Theta(x_2)
\over \sinh \epsilon (x_1-x_2)})_{x_2}
$$
\begin{equation}
{}\label{Z4}
\end{equation}
$$
X^2=-{1\over 4\epsilon^2}({\sinh \epsilon (x_1+2x_2-N)\over \sinh \epsilon
(x_1-x_2)})_{x_1}-(\coth \epsilon (x_1-x_2)\bar \Theta(x_2)+{\Theta(x_1)
\over \sinh \epsilon (x_1-x_2)})_{x_1}
$$
This formulae are the result of the exact integration of
the system (\ref{ZZZ}).  The remaining additional conditions
follow from the corresponding  equations  for  invariant  (in  the
sense of independence of the momentum $M$) structural functions.
They may be summarized as
\begin{equation}
(\ln X^1)_{x_2}+(\ln X^2)_{x_1}=0\label{ADC}
\end{equation}
To satisfy them it is natural to make a special choice of
arbitrary functions $\Theta,\bar \Theta$ in the general
solution (\ref{Z4}).

After performing all operations of the differentiation in (\ref{Z4})
we obtain for $X^1$
$$
X^1={1\over 4\epsilon}({\sinh \epsilon (3x_1-N)\over \sinh^2 \epsilon
(x_1-x_2)})+{\Theta(x_1)\over \sinh^2 \epsilon (x_1-x_2)}+{\sinh \epsilon
(x_1-x_2)\bar \Theta(x_2)_{x_2}+\cosh \epsilon (x_1-x_2)\bar \Theta(x_2)
\over \sinh^2 \epsilon (x_1-x_2)}
$$
If we want to satisfy additional condition (\ref{ADC}), it is natural
to find the solution in the form
$X^{1,2}={\psi (x_{1,2})\over \sinh^2 \epsilon (x_1-x_2)}$.
Under such kind of assumption for functions $\Theta, \bar \Theta$ we uniquely
obtain $\Theta=A\cosh \epsilon x +B\sinh \epsilon x,\bar \Theta=C\cosh
\epsilon x +D\sinh \epsilon x$ where $A,B,C,D$ are arbitrary numerical
constants.  Finally, for the function $\psi$ which parameterizes the solution
of the system (\ref{Z4}) we obtain
$$
\psi(x)={1\over 4\epsilon}\sinh \epsilon (3x-N) +(A+C)\sinh \epsilon x+(B+D)
\cosh \epsilon x
$$
The last  expression  for  $\psi$  may  be  parameterized  in  the
following form:
$$
\psi(x)={1\over \epsilon}\sinh \epsilon (x-N_1)\sinh \epsilon (x-N_2)
\sinh \epsilon (x-N_3)
$$
as one can see representing $\sinh$ as the difference of the exponents
with further multiplication term by term.

\subsection{The quantum group case}

Now in the realization (\ref{7}) we consider $(m,l_1,l_2)$ as coordinates
and $(M,L_1,L_2)$ as momenta of the three independent (mutually comutative)
Heisenberg algebras, and $(N_1,N_1,N_2)$ as "cyclic" variables that
commutate with all other generators involved.
We  choose  $f=a\sinh \epsilon x$, since the quantum group
must have as its classical limit the functional group of
the previous subsection.

Commutation relations $\left[X^+_1,X^-_2\right]=0,\quad \left[X^+_2,
X^-_1\right]=0$ allow us to reconstruct the dependence of structural functions
$f^{1,2},\bar f^{1,2}$ upon the momentum $M$ exactly in the form
(\ref{SF}).

The only unused up to now commutation relation
$$
\left[X^+_2,X^-_2\right]=a\sinh \epsilon (M-2(L_1+L_2)+N_1+N_2+N_3)
$$
together with (\ref{7}) has as its consequence an additional system
of equations determining the structural functions:
$$
f^1(L_1,L_2-1) \bar f^2(L_1-1,L_2)=f^1(L_1,L_2+1) \bar f^2(L_1+1,L_2),
$$
\begin{equation}
{}\label{AADD}
\end{equation}
$$
\bar f^1(L_1,L_2-1) f^2(L_1-1,L_2)=\bar f^1(L_1,L_2+1) f^2(L_1+1,L_2)
$$
Finally, we will present the last remaining equation using the following 
notations, $X^1_{\pm}\equiv -f^1(L_1\pm 1,L_2) \bar f^1(L_1\pm 1,L_2),$ $
X^2_{\pm}\equiv f^2(L_1,L_2\pm 1) \bar f^2(L_1,L_2\pm 1)$ for new unknown
functions and $x_1\equiv L_1,x_2\equiv L_2$ for new coordinates.
The equation takes on the form:
$$
  -\sinh \epsilon(x_1-M-1) X^1_- +\sinh \epsilon(x_1-M+1) X^1_+
  -\sinh \epsilon (M-L_2+1) X^2_-+\sinh \epsilon(M-L_2-1) X^2_+=
$$
\begin{equation}
  2a\sinh \epsilon (M-2(L_1+L_2)+N_1+N_2+N_3)   \label{FE}
\end{equation}

After resolution of the last equation (equating terms with
$\sinh M,\cosh M$ on both sides) we come to a system of
two equation for determining the two unknown functions $X^{1,2}$:
\begin{eqnarray}
-e^{\epsilon (x_1-1)} X^1_-+e^{\epsilon (x_1+1)} X^1_+-e^{\epsilon (x_2-
1)} X^2_-+e^{\epsilon (x_2+1)} X^2_+ &=& 2e^{\epsilon (2x_1+2x_2-N)}
                                                        \nonumber  \\
-e^{-\epsilon (x_1-1)} X^1_-+e^{-\epsilon (x_1+1)} X^1_+-e^{-\epsilon
(x_2-1)} X^2_-+e^{-\epsilon (x_2+1)} X^2_+ &=& 2e^{-\epsilon (2x_1+2x_2+N)}
                                                        \label{FE2}
\end{eqnarray}
where $N=N_1+N_2+N_3$.

To solve the last system we introduce the operation
of discrete differentiation as
$$
\Delta_i F(x_1,...x_n)\equiv {F(x_1,.x_i+1,..x_n)-F(x_1,.x_i-1,..x_n)\over 2}
$$
These operations mutually commute  $\Delta_i\Delta_j=
\Delta_j\Delta_i$, and posses the linearity properties in the
following sense
$$
\Delta_i (F^1+F^2)=\Delta_i F^1+\Delta_i F^2,\quad \Delta_i (CF)=C\Delta_i F
$$
if the function $C$ is independent of the $x_i$ coordinate.

The system (\ref{FE}),(\ref{FE2}) may be rewritten in terms of these operations as
$$
\Delta_1(e^{\epsilon x_1} X^1)+\Delta_2(e^{\epsilon x_2} X^2)=
e^{\epsilon (2x_1+2x_2-N)}
$$
$$
\Delta_1(e^{-\epsilon x_1} X^1)+\Delta_2(e^{-\epsilon x_2} X^2)=
e^{-\epsilon (2x_1+2x_2-N)}
$$

As in previous subsection, we seek for a solution of the last system
in the form
$$
X^1=\Delta_2 (Y^2),\quad X^2=\Delta_1 (Y^1)
$$
Keeping in mind the relation (which can be easily checked)
$$
\Delta_2 \Delta_1 e^{\epsilon (2x_1+2x_2)}=(\sinh 2\epsilon)^2 e^{\epsilon
(2x_1+2x_2)}
$$
we obtain the linear system of algebraic equations
determining $Y^1,Y^2$.
After resolution of this system, the solution of the initial system
takes the form
$$
X^1=\Delta_2 ({1\over \sinh^2 2\epsilon}{\sinh \epsilon (2x_1+x_2-N)\over
\sinh \epsilon (x_1-x_2)}+\coth \epsilon (x_1-x_2)\Theta(x_1)+{\bar
\Theta(x_2)
\over \sinh \epsilon (x_1-x_2)})
$$
\begin{equation}
{}\label{Z5}
\end{equation}
$$
X^2=-\Delta_1 ({1\over \sinh^2 2\epsilon}\,
    {\sinh \epsilon (x_1+2x_2-N)\over \sinh \epsilon (x_1-x_2)}
     +\coth \epsilon (x_1-x_2)\bar
     \Theta(x_2)+{\Theta(x_1)
    \over \sinh \epsilon (x_1-x_2)})
$$
After all necessary calculations we obtain from (\ref{Z5}) the finally
expression:
\begin{eqnarray}
  X^1={1\over 2\sinh 2\epsilon}
      {\sinh \epsilon (3x_1-N)\over
      \sinh \epsilon (x_1-x_2-1)\sinh \epsilon (x_1-x_2+1)} +
  \nonumber \\
 {\sinh 2 \epsilon \Theta(x_1) +
  \bar \Theta(x_2+1) \sinh \epsilon (x_1-x_2+1)-
  \bar \Theta(x_2-1)\sinh \epsilon (x_1-x_2-1)
  \over \sinh \epsilon (x_1-x_2-1)\sinh \epsilon (x_1-x_2+1)}
  \label{1eq}
\end{eqnarray}
and the same kind of expression for $X^2$.
If we want to satisfy the additional
condition (\ref{AADD}) which can be rewritten in the form
$$
X^1(L_1,L_2-1)X^2(L_1-1,L_2)=X^1(L_1,L_2+1)X^2(L_1+1,L_2)
$$
then it is necessary to find a solution as
\begin{eqnarray}
  X^1={\psi_1 (x_1)\over \phi(x_1-x_2)},\quad X^2={\psi_2 (x_2)\over
  \phi(x_1-x_2)}
  \label{2eq}
\end{eqnarray}

It is easy to see that with this form of solution the additional condition
(\ref{AADD}) is satisfied automatically.
Comparing Eqs.\ (\ref{1eq}) and (\ref{2eq}) we come to the unique
possible choice of the function $(\Theta,\bar \Theta)$,
$$
\Theta(x)=A\sinh x+B\cosh x,\quad \bar \Theta(x)=C\sinh x+D\cosh x
$$
And as consequence we obtain $\psi_1(x)=\psi_2(x)=\sinh \epsilon
(3x-N)+p\sinh \epsilon x +q\cosh \epsilon x \equiv \sinh \epsilon (x-N_1)
\sinh \epsilon (x-N_2) \sinh \epsilon (x-N_3)$. Finally, the solution of
the initial system satisfying all necessary additional conditions 
may be written in the form:
$$
X^1={\sinh \epsilon (x_1-N_1) \sinh \epsilon (x_1-N_2) \sinh \epsilon (x_1-N_
3)\over 2 \sinh 2\epsilon \sinh \epsilon (x_1-x_2-1)\sinh \epsilon 
(x_1-x_2+1)},
$$
\begin{equation}
{} \label{FF}
\end{equation}
$$
X^2={\sinh \epsilon (x_2-N_1) \sinh \epsilon (x_2-N_2) \sinh \epsilon (x_2-N_
3)\over 2\sinh 2\epsilon \sinh \epsilon (x_1-x_2-1)\sinh \epsilon 
(x_1-x_2+1)}
$$

\section{General case of arbitrary $n$}

\subsection{The algebra representation level}

Let us assume that the generators of the simple roots and Cartan elements 
of $U(n+1)$ algebra may be presented in the form
\begin{equation}
X^+_s=\sum_{k=1}^s e^{l^s_k} g^s_k e^{l^s_k},\quad X^-_s=\sum_{k=1}^s
e^{-l^s_k} \bar g^s_k e^{-l^s_k} \label{20}
\end{equation}
$$
h_s=-{1\over 2}\sum_{r=1}^{s-1} L^{s-1}_r+\sum_{k=1}^s L^s_k-{1\over 2}
\sum_{l=1}^{s+1} L^{s+1}_l\quad 1\leq s \leq n
$$
where different from zero commutators of the operators involved are 
$$
 [L^s_k,L^t_l]=\delta_{st} \delta_{kl}I
$$
We assume that the structural functions are the "factorizable" 
and as functions of their arguments may be represented in the 
following form:
$$
g^s_k=F^s_k(L^{s+1},L^s)f^s_k(L^{s-1},L^{s-1}) \quad \bar g^s_k=
\bar F^s_k(L^{s+1},L^s)\bar f^s_k(L^{s-1},L^{s-1})
$$
The reader without any difficulty may identify $L^1$ with $M$, $L^2$ with
$L_1,L_2$ and $L^3$ with $N_1,N_2,N_3$ from the previous section.

We assume also that all necessary commutation relations for $1\leq s \leq
(n-1)$ are correctly satisfied and:
$$
(F^{n-1}_k)^2=(\bar F^{n-1}_k)^2={\prod_{r=1}^n \sinh \epsilon (L^{n-1}_k-
L^n_r)\over \Phi (L^{n-1})}
$$
where function $\Phi$ is translation invariant with respect to the 
simultaneous shift of all of the arguments $L^{n-1}_k$.
We to prove by induction that the squares of the structural 
functions $F^n_i=\bar F^n_i$ conserve their form and find 
the explicit expressions for the denominator.

It is obvious that under the above restrictions the commutation relations
between generators of Cartan subalgebra $h_l$ and generators of the simple
roots $ X^{\pm}_k $ are correctly satisfied. It is also clear 
that the generators
$ X^{\pm}_n $ commute with all generators $X^{\mp}_k$ with
$1\leq k \leq (n-2) $ because they act on essential different arguments.
And at last,  the commutation relations 
$$
  [X^{\pm}_n,X^{\mp}_{n-1}]=0
$$
allow us to reconstruct the explicit dependence of structural functions
$f^n_k,\bar f^n_k$ on their arguments $L^{n-1}_k$. 

As a direct corollary of the last commutation relations we have:
$$
f^n_k(L^n;..,L^{n-1}_r-1,..)\bar F^{n-1}_r(..,L^n_k-1,..;L^{n-1})=
f^n_k(L^n;..,L^{n-1}_r+1,..)\bar F^{n-1}_r(..,L^n_k+1,..;L^{n-1})
$$
$$
\bar f^n_k(L^n;..,L^{n-1}_r-1,..) F^{n-1}_r(..,L^n_k-1,..;L^{n-1})=
\bar f^n_k(L^n;..,L^{n-1}_r+1,..) F^{n-1}_r(..,L^n_k+1,..;L^{n-1})
$$
which must be satisfied for all the possible choices of $k$ and $r$ indexes.
Keeping in mind the explicit form of the structural functions 
$F^{n-1}_k=\bar F^{n-1}_k$ as assumed above, we can to
resolve the last equations in the form:
\begin{equation}
g^n_k=F^n_k(L^{n+1};L^n)\sqrt {\prod^{n-1}_{r=1} \sinh \epsilon (L^n_k-
L^{n-1}_r)},\quad
\bar g^n_k=\bar F^n_k(L^{s+1};L^s)\sqrt {\prod^{s-1}_{r=1} \sinh \epsilon
(L^n_k-L^{n-1}_r)}
\label{21}
\end{equation}

The only commutation relation which has not satisfied up to now, is 
\begin{equation}
 [X^+_n,X^-_n]=\sinh \epsilon(h_n) \label{22}
\end{equation}

It consists of "diagonal" part (which doesn't contain the
coordinates of Heisenberg subalgebras $l_i$) and nondiagonal one
(in above sense).
The arised equation for the diagonal part can be naturally 
written using notations: 
$$
X^{\pm n}_k=F^n_k(L^{n+1};..L^n_k \pm 1,..)\bar F^n_k(L^{n+1};..,L^n_k \pm
1...)
$$
\begin{equation}
\sum_{k=1}^n \prod^{n-1}_{r=1} {\sinh \epsilon (L^n_k+1-L^{n-1}_r)} X^{+n}_k-
\sum_{k=1}^n{\prod^{n-1}_{r=1} \sinh \epsilon (L^n_k-1-L^{n-1}_r)} X^{-n}_k=
\label{23}
\end{equation}
$$
\sinh\epsilon(\sum_{r=1}^{n-1} L^{n-1}_r+2\sum_{k=1}^n L^n_k-\sum_{j=1}^{n+1}
L^{n+1}_j)
$$
The unknown functions  $X^{\pm n}_k$ depend only on $L^{n+1},L^n$ variables 
(as it follows from their definition) and thus, with the respect 
to the variables $L^{n-1}$ (\ref{23}) they must be satisfied identically. 
Let us consider the structure of the product in left-hand side
of (\ref{23})  representing each hyperbolic function as a sum of 
two exponents:
$$
\prod^{n-1}_{r=1} {\sinh \epsilon (x-L^{n-1}_r)}={1\over 2^{n-1}} \sum_{k=0}
^{n-1} A_k e^{\epsilon [(n-1)-2k] x]}
$$
It is clear that that $A_0=e^{-\epsilon \sum_{r=1}^{n-1} L^{n-1}_r},$
$A_{n-1}=(-1)^{n-1}e^{\epsilon \sum_{r=1}^{n-1} L^{n-1}_r}$ 
and all other $A_k$ are some complicate symmetrical functions 
constructed from the components of the $(n-1)$-dimensional vector
$L^{n-1}$. In the right-hand side of (\ref{23}) (after the 
decomposition of the hyperbolic function) the dependence on $L^{n-1}$ is
concentrated only in the two exponents. Thus equating the coefficients 
at such terms in both sides of (\ref{23}) we come to the system of $n$
equations for $X^ n_k$ functions (we introduce notations
$L^{n-1}_k\equiv x_k$ and for a time put $\epsilon=1$):
$$
\sum_{k=1}^n [e^{(n-1)(x_k+1)} X^{+n}_k-e^{(n-1)(x_k-1)} X^{-n}_k]=
2^{n-2} e^{2\sum_{k=1}^n x_k-\sum_{j=1}^{s+1} L^{s+1}_j}
$$
\begin{equation}
\sum_{k=1}^n [e^{s (x_k+1)} X^{+s}_k-e^{s (x_k+1)}X^{-s}_k]=0,\quad
s\equiv n-1-2r,\quad r=1,2,..(n-2) \label{24}
\end{equation}
$$
\sum_{k=1}^n [e^{-(n-1)(x_k+1)} X^{+n}_k-e^{-(n-1)(x_k-1)} X^{-n}_k]=
(-1)^n 2^{n-2} e^{-2\sum_{k=1}^n x_k+\sum_{j=1}^{s+1} L^{s+1}_j}
$$

The zero value of "nondiagonal" part of (\ref{22}) is equivalent to
additional conditions which structural functions $F,\bar F$ must satisfy:
$$
F^n_k(L^{n+1};..,L^n_j-1,..) \bar F^n_j(L^{n+1};..,L^n_k-1,..)=
F^n_k(L^{n+1};..,L^n_j+1,..)\bar F^n_j(L^{n+1};..,L^n_k+1,..)
$$
\begin{equation}
{}\label{25}
\end{equation}
$$
\bar F^n_k(L^{n+1};..,L^n_j-1,..) F^n_j(L^{n+1};..,L^n_k-1,..)=
\bar F^n_k(L^{n+1};..,L^n_j+1,..) F^n_j(L^{n+1};..,L^n_k+1,..)
$$
{}From (\ref{25}) we see that the solution $F^n_j=\bar F^n_j$ is
the possible one and functions $X^n_k\equiv X^k $ being the solution
of (\ref{23})), must satisfy additional conditions:
\begin{equation}
X^k(L^{n+1};..,L^n_j-1,..) X^j(L^{n+1};..,L^n_k-1,..)=
X^k(L^{n+1};..,L^n_j+1,..) X^j(L^{n+1};..,L^s_n+1,..)\label{26}
\end{equation}
The reader can easily obtain from general equations of the
present subsection  all results of the previous
section for the case $n=2$. At this point we interrupt
our consideration for a moment to represent  general  solution  of 
continuous
version of equations (\ref{24}).

\subsection{General solution of linear system in the continuous limit}

In this subsection we give a general solution of the system (\ref{24})
in continuous limit or at the level of the functional group approach. 
In this limit the system (\ref{24}) may be obviously rewritten as: 
$$
\sum_{k=1}^n (e^{(n-1) x_k} X^k)_{x_k}=-2^{n-2} e^{2\sum_{k=1}^n x_k-L_{n+1}}
$$
\begin{equation}
\sum_{k=1}^n (e^{s x_k} X^k)_{x_k}=0,\quad
s\equiv n-1-2r,\quad r=1,2,..(n-2) \label{25a}
\end{equation}
$$
\sum_{k=1}^n (e^{-(n-1)x_k} X^k)_{x_k}=(-1)^{n-1}2^{n-2} e^{-2\sum_{k=1}^n x_k
+L_{n+1}}
$$
(compare with the the corresponding equation (\ref{ZZZ}) of the 
previous section).
We have introduce the following abbreviation, $\sum_{j=1}^{n+1} L^{n+1}_j
\equiv L_{n+1}$.

Keeping in mind the known for us ways of resolution of the systems of 
such kind \cite{Leznov} we find the solution in the form:
$$
X^i=(Y^i)_{x_1,....x_{i-1},x_{i+1},..x_n}
$$
After substitution of these expressions in (\ref{25a}) we come to the
linear system of algebraic equations with the known right-hand side 
(we have used the obvious relation
$(e^{2\sum_{k=1}^n x_k-L_{n+1}})_{x_1,...x_n}=
 2^n e^{2\sum_{k=1}^n x_k-L_{n+1}})$,
to determine the unknown functions $Y^i$: 
$$
\sum_{k=1}^n  e^{(n-1) x_k} Y^k=2^{-3} e^{2\sum_{k=1}^n x_k-L_{n+1}}+
\Phi^{n-1}
$$
\begin{equation}
\sum_{k=1}^n e^{s x_k} Y^k=\Phi^{s},\quad s\equiv n-1-2r,\quad r=1,2,..,(n-2)
\label{28}
\end{equation}
$$
\sum_{k=1}^n e^{-(n-1)x_k} Y^k=2^{-3} e^{-2\sum_{k=1}^n x_k+L_{n+1}}+
\Phi^{-n+1}
$$
All functions $\Phi^{n-1-2s},\quad 0\leq s \leq (n-1)$ are solutions of the
single equation $\Phi_{x_1,....x_n}=0$ (differentiation with the respect to
all arguments of the problem!).

The following notations will be useful for resolution of (\ref{28}). Let
$W_r$ be the coefficient of the polynomial of $(n-1)$ degree expressed 
by means its roots:
$$
P_{n-1}(z)=W_0 z^{n-1}+W_1 z^{n-2}+....+W_{n-1}=\prod_{k=1}^{n-1} (z-z_k),
\quad (W_0=1)
$$
We denote the coefficients of the polynomial as $W^{(k}_r$; the $(n-1)$ roots
of this polynomial coincide with the $(n-1)$ exponents $e^{2x_i}$ 
except of only one $e^{2x_k}$.

In this notations resolution of the algebraic system of equations 
(\ref{28}) may be represented as:
$$
Y^k={1\over 2^2}{(\sinh) \cosh (2x_k+\sum^{'k}_{i=1} x_i-L_{n+1})\over
\prod^{'n}_{i=1} \sinh (x_k-x_i)}+{\sum_{s=0}^{n-1} \Phi^{(n-1-2s} W^{(k}_s
\over \prod^{'n}_{i=1} \sinh (x_k-x_i)}
$$
In the numenator of the last expression the sign of $\cosh$ arises 
in the case of odd $n=2r+1$, and the sign $\sinh$ -- in the case of 
even $n=2r$, and  $L_{n+1}=\sum_{j=1}^{n+1} L^{n+1}_j$.
By consequent differentiation of $Y^k$ with respect to all independent 
arguments, except of $x_k$, we come to the explicit solution of the 
initial system (\ref{28}):
\begin{equation}
X^k={1\over 2^2}{(\sinh) \cosh ((n+1)x_k-L_{n+1})\over \prod^{'n}_{i=1}
\sinh^2 (x_k-x_i)}+({\sum_{s=0}^{n-1} \Phi^{(n-1-2s} W^{(k}_s
\over \prod^{'n}_{i=1} \sinh^2 (x_k-x_i)})_{x_1,..x_{k-1}x_{k+1},..x_n}
\label{29}
\end{equation}
The first term in (\ref{29}) is a particular solution of
the inhomogeneous system (\ref{28}), the second one is a general solution 
of the homogeneous part of it. Of course, after performing all of the 
differentiations, the general solution will contain only one 
single function which satisfies the scalar equation 
$(\Phi)_{x_1,...x_n}=0$. 

Below we demonstrate only a particular solution of the homogeneous system,
sufficient for the aim to satisfy all necessary additional conditions of our
problem.
We will directly show that:
\begin{equation}
X^i={e^{[(n+1)-2s]x_i}\over \prod^{'n}_{k=1} \sinh^2 (x_i-x_k)}=
(e^{[(n+1)-2s]x_i}\prod^{'n}_{k=1} {e^{2 x_i}+e^{2 x_k}\over e^{2 x_i}-
e^{2 x_k}})_{x_1,..x_{i-1}x_{i+1},..x_n},\quad 2\leq s \leq (n-1) \label{BS}
\end{equation}
In the transformation above we have used the equality 
$(\coth (x_i-x_k))_{x_k}=\sinh^{-2}
(x_i-x_k)$ and the decomposition of $\sinh,\cosh$ functions into 
two exponents.
Replacing $X^i$ in such a form in the left-hand side of the system
(\ref{28}) we have:
\begin{equation}
(\sum_{k=1}^n  e^{[(n+1)-2s+(n-1)-2r]x_k} \prod^{'n}_{i=1} {e^{2 x_k}+
e^{2 x_i}\over e^{2 x_k}-e^{2 x_i}})_{x_1,..x_n},\quad 0\leq r \leq (n-1)
\label{30}
\end{equation}
In all cases from restrictions on $s,r$ we have $-n+2\leq n-r-s \leq n-2$.
In the further transformations of the last expression is suitable to 
distinguish between two different cases: $0\leq n-r-s \leq n-2$ and $-n+2\leq
n-r-s
\leq 0$.
In the first one we introduce $\lambda_i\equiv e^{2x_i}$ and rewrite
(\ref{30}) in the form
$$
2^n(\lambda_1,...,\lambda_n)(\sum_{k=1}^n  \lambda_k^{(n-s-r)}
{\prod^{'n}_{i=1} (\lambda_k+\lambda_i)\over \prod^{'n}_{i=1}
(\lambda_k-\lambda_i)})_{\lambda_1,..\lambda_n},\quad 0\leq n-r-s \leq n-2
$$
and in the second case we will get the same expression 
(up to unessential factors) introducing $\lambda_i\equiv e^{-2x_i}$.

In both cases we have a symmetrical function under the sign of 
differentiation. 
Reducing to a common denominator we obtain the ratio of two $n$ 
dimensional polynomial functions, one of which (denominator) is exactly 
the Wandermond determinant (the single function which is antisymmetrical 
with respect to permutation of each pair of coordinates). 
Thus, the nominator must also be an antisymmetrical polynomial 
(since the ratio is a symmetrical one!)  This is impossible if
$ n-r-s$ is less than $n-1$ (the degree of nominator in this case is less
than the degree of Wandermond determinant).

We thus prove that (\ref{BS}) is a particular solution of homogeneous
system (\ref{28}).

Adding the particular solution (\ref{BS}) to the particular solution of
inhomogeneous ones (\ref{29}) we come to the following solution of
inhomogeneous system (\ref{28}) which satisfies all necessary additional
conditions as one can check directly:
\begin{equation}
X^k={1\over 2^2}{(\sinh) \cosh ((n+1)x_k-L_{n+1})+\sum_{s=1}^{n-2}
(A_s e^{((n+1)x_k-2s)}+B_se^{(-(n+1)x_k+2s)}\over \prod^{'n}_{i=1}
\sinh^2 (x_k-x_i)}\equiv \label{EF}
\end{equation}
$$
{1\over 2^2}{\prod^{n+1}_{j=1}\sinh (x_k-L^{n+1}_j)\over
\prod^{'n}_{i=1} \sinh^2 (x_k-x_i)}
$$
One can be convinced in the possibility of the last representation
by decomposition of $\sinh$ into two exponents in each factor of 
the product with further multiplication by terms.

The last representation for $X^n$ proves the 
accomplish the induction procedure in the case of the founctional group.

\subsection{Solution of the problem in the discrete case}

In this subsection we present the solution of the finite difference system
(\ref{24}) together with all additional conditions (\ref{26}) and prove by
this way the induction assumption.

As the reader will see, the most surprising feature of such an approach  
is that both the calculations themselves and the final result
are not changed essentially in comparison with the continuous 
functional group case.

Being rewritten in terms of the operation of discrete differentiation (see 
subsection 3 of the previous section) system (\ref{24}) takes the form
$$
\sum_{k=1}^n \Delta_k (e^{(n-1)x_k}X^k)=-2^{n-3} e^{2\sum_{i=1}^n x_i-L_
{n+1}}
$$
\begin{equation}
\sum_{k=1}^n \Delta_k (e^{sx_k}X^k)=0,\quad s=\equiv (n-1-2r),\quad r=1,2,..
(n-2) \label{BB}
\end{equation}
$$
\sum_{k=1}^n \Delta_k (e^{-(n-1)x_k}X^k)=(-1)^n 2^{n-3} e^{-2\sum_{i=1}^n x_i+
L_{n+1}}
$$
Let us seek for the solution of the last system in the usual for us form:
$$
X^{k}=(Y^k)_{x_1,...x_{k-1}x_{k+1}...x_n}
$$
Replacin this expression in (\ref{BB}) and using the
obvious equality $\Delta_i e^{\pm 2x_i}=\pm \sinh 2 e^{\pm 2x_i}$ we
come to linear algebraic equations for determining unknown functions $Y^k$:
$$
\sum_{k=1}^n  e^{(n-1) x_k} Y^k=2^{n-3}(\sinh 2)^{-n} e^{2\sum_{k=1}^n x_k-
L_{n+1}}+\Phi^{n-1}
$$
\begin{equation}
\sum_{k=1}^n e^{s x_k} Y^k=\Phi^{s},\quad s\equiv n-1-2r,\quad r=1,2,..,(n-2)
\label{BBB}
\end{equation}
$$
\sum_{k=1}^n e^{-(n-1)x_k} Y^k=-2^{n-3} (\sinh 2)^{-n} e^{-2\sum_{k=1}^n x_k+
L_{n+1}}+\Phi^{-n+1}
$$
where all $\Phi^a$ are solutions of the same single equation
$$
(\Delta_1,\Delta_2,....\Delta_n) \Phi^a=0
$$
In the last expression the discrete differentiation is performed 
with respect to all coordinates of the problem.

The system (\ref{BBB}) coincides up to the unessential factors with
(\ref{28}).
Thus, its general solution (up to the obvious corrections) is the same as
(\ref{29}) (of course with discrete differentiation).

In performing the last operation the following equality is necessary
$$
\Delta_i {(\cosh)\sinh (x_i+A)\over \sinh (x_i-x_k)}={\sinh 2\over 2}
{(\cosh)\sinh (x_k+A)\over \sinh (x_i-x_k-1)\sinh (x_i-x_k+1)}
$$
Finally for the solution of (\ref{BB}) we obtain: 
\begin{equation}
X^k={1\over 2 \sinh 2}{(\sinh) \cosh ((n+1)x_k-L_{n+1})\over \prod^{'n}_{i=1}
\sinh (x_k-x_i-1)\sinh (x_k-x_i+1)}+X^k_0
\label{NNB}
\end{equation}
where $X^k_0$ is general solution of the homogeneous system.

As  the reader may notice, the only difference  compared  to the 
continuous case is the appearence of  the product of $\sinh$ functions 
with shifted arguments in denominator  at the  place of the single square. 

As in the continuous case it is not difficult to verify that the form
$X^k={\Psi (x_k)\over \Phi (x_k-x_i)}$ is selfconsistent with all additional
restrictions. We want to show now that the corresponding solution of
a homogeneous system (the second term in (\ref{NNB})) may be chosen.

By the same way and methods we show that the homogeneous system
possesses the solutions of the form
$$
 X^i={e^{[(n+1)-2s]x_i}\over \prod^{'n}_{k=1} \sinh (x_i-x_k-1)\sinh
 (x_i-x_k+1)}=
 \Delta_1... \Delta_{i-1}\Delta_{i+1}....
 \Delta_n ( e^{[(n+1)-2s]x_i}\prod^{'n}_{k=1}
  { e^{2 x_i} +e^{2 x_k} \over  e^{2 x_i}- e^{2 x_k}}),
$$
\begin{equation}
2\leq s \leq (n-1) \label{BSS}
\end{equation}

Taking the sum of this solution and the particular solution of the
inhomogeneous system (\ref{NNB}) we obtain  $X^k$ in the factorized form,
$$
X^k={1\over 2\sinh 2}{\prod^{n+1}_{j=1}\sinh (x_k-L^{n+1}_j)\over
\prod^{'n}_{i=1} \sinh (x_k-x_i-1)\sinh (x_k-x_i+1)}
$$
The last expression accomplishes the prove of the induction procedure in
the quantum algebra case.

\section{Outlook}

The  main  result  of  the  present  paper  consists  in  the explicit
realization of irreducible representations of quantum algebras
$A^q_n$ of the semisimple serie.  The method used by us is essentially
different from the comonnly  used approaches to the solution
of this problem in literature.
  Among the fundamental suggestions we use only the
Gel'fand-Zeitlin selection rules and commutation relations
among generators of the simple roots and Cartan elements of the $A^q_n$
algebra with arbitrary functional dependence in principal commutation
relation  $[X^+_i,X^-_j]=\delta_{ij} \phi_i(h_i)$.

Already for the case of $A^q_2$ algebra (and for all $A^q_n,\, n>2$)
such construction is selfconsistant only in the case of the
the choice of $\phi_i$ functions in the standard trigonometrical form.
No additional assumption about the existence of a co-product
is made. In fact, the result can be obtained at the level of the functional
group (classic limit) and the only thing which is necessary is to check
it at the algebra-theoretical level.

{}From the physical point of view this means that the quasi-classical approach
in fact leads to a correct quantum result.  We are in a position to describe
this situation but would not like to make final conclusions at this step.

\section{Acknowledgements}

Author is indebted to the Instituto de Investigaciones en Matem\'aticas
Aplicadas y en Sistemas, UNAM and especially  to  its  director
Dr.\  I.\  Herrera  for beautiful conditions for his work.
Permanent company and discussions with N.\ Atakishiyev, S.M. Chumakov,
K.B.\ Wolf and P.\ Winternitz allow to the author to finish this
paper in the shortest time.

\section{Appendix I}

Choosing the new unknown function sas $F^1(X)\equiv X^2_x,F^2(Y)=Y^2_y$
and $x_1=X^2,x_2=Y^2$ as new independent coordinates,
we rewrite (\ref{ZZ}) in the form of linear equation:
$$
(x_1 F^1_{x_1}+x_2 F^2_{x_2})(x_1-x_2)=(F^1-F^2)(x_1+x_2)
$$
It possesses the three obvious solutions $F=x,F=1,F={1\over x}$. Their
linear combination with arbitrary coefficients is also a solution.
This solution is the general one. Indeed,  differentiating twice
the last equation with respect, say, to $x_1$ coordinate we come to
a system of selfconsistent equations of the form presented above.



\begin{thebibliography}{20}

\bibitem{Leznov} A.N.\ Leznov, {\it To the Gel'fand-Tsetlin Realization of
Irreducible Representations of Classical Semisimple Algebras,}
Preprint of IIMAS -- UNAM, No. 77,
Febrero, 1998.

\bibitem{Tsetlin} I.M.\ Gel'fand and M.L.\ Tsetlin
Dokl. Akad. Nauk SSSR {\bf 71}, 825-828 (1950);
Dokl. Akad. Nauk SSSR {\bf 71}, 1017-1020 (1950).

\bibitem{Eisenhart} L.P.\ Eisenhart, {\it Continuous Groups of
Transformations}, Princeton N.J.: Princeton University Press, 1933.

\bibitem{Malkin} A.N.\ Leznov, I.A.\ Malkin and V.I.\ Man'ko
  {\it Canonical transformations and representation theory of Lie groups.}
Trudy FIAN {\bf 96}, 24-72 (1977).


\end{thebibliography}
\end{document}